\documentclass[11pt]{article}

\usepackage[margin=1in]{geometry}
\usepackage{amsmath,amssymb,amsthm,mathtools}   
\usepackage{hyperref}
\usepackage{enumitem,xspace}

\newtheorem{theorem}{Theorem}
\newtheorem{assumption}{Assumption}
\newtheorem{lemma}{Lemma}

\DeclareMathOperator{\ALG}{ALG}
\DeclareMathOperator{\OPT}{OPT}

\newcommand{\R}{\mathbb R}

\newcommand{\bO}{\mathbf{O}}
\newcommand{\bL}{\mathbf{L}}
\newcommand{\bx}{\mathbf{x}}
\newcommand{\by}{\mathbf{y}}

\newcommand{\bzero}{\mathbf{0}}
\newcommand{\bell}{\boldsymbol{\ell}}
\newcommand{\bpi}{\boldsymbol{\pi}}

\title{Tight Analysis for Online Generalized Network Design\\ with Power
Functions}
\author{Karl Li\thanks{Department of Mathematics, University of Michigan, Ann
Arbor, USA
(\texttt{kkli@umich.edu}, \texttt{nuvole@umich.edu}).}
\and Yanzhi Li\footnotemark[1]
\and Viswanath Nagarajan\thanks{
Department of Industrial and Operations Engineering, University of Michigan,
Ann Arbor, USA
(\texttt{viswa@umich.edu}).
Research supported in part by NSF grant CCF-2418495.
}
}

\date{}

\begin{document}
\maketitle

\begin{abstract}
  We study the online generalized network design problem with heterogeneous
  power costs. Here, a sequence of $N$ requests arrive online, each of which
  needs to be assigned to some subset of available resources $E$. Each
  resource $e\in E$ has its own coefficient $\xi_e>0$ and exponent $p_e\ge 1$,
  where the cost of a load vector $L\in \mathbb{R}^E_+$ is $
  \Phi(L)=\sum_{e\in E}\xi_e L_e^{p_e}$. We provide a simple analysis for the
  natural greedy algorithm that $\tau$-approximately minimizes the increase in
  $\Phi$ for each online request. Our competitive ratio is at most
  $(\frac{p}{\ln 2})^p$ when $\tau=1$ and is roughly $(\tau \, p)^p$ as $\tau$
  grows large; here $p=\max_{e\in E} p_e$. This improves over a prior
  competitive ratio of $ 2(\mathsf e\tau p)^p$ that relied on a modified
  greedy algorithm and a more complex analysis. Moreover, our bound (with
  $\tau=1$) is asymptotically optimal: it matches the $(1.443p)^p$ lower bound
  for deterministic online algorithms. We also provide a nuanced
  exponent-dependent bound that, combined with prior results, provides an
  improved competitive ratio for (dis)economies of scale cost functions; here,
  the cost of resource $e$ has the form $\sigma_e + \xi_e\cdot x^{p_e}$ if
  $x>0$ (and zero if $x=0$). These results apply to several fundamental
  network problems such as multicommodity routing (where $\tau=1$), Steiner
  tree connectivity (where $\tau=1.39$) and directed Steiner connectivity
  (where $\tau$ is poly-logarithmic).
\end{abstract}

\section{Introduction.}

Online resource-allocation and network-design problems often involve
heterogeneous resources
(servers, links and machines).
In energy-aware systems, the cost incurred by a resource having load $x$ is
modeled by
a non-linear convex power function $x \mapsto \xi_e x^{p_e}$, where both the
coefficient $\xi_e>0$ and the
exponent $p_e\ge 1$ may depend on the resource $e$.
The exponent $p_e$ typically lies between 1 and 3
\cite{AndrewsAZZ12,WiermanAT12}.

While classical work such as \cite{AwerbuchAGKKV95} addresses load balancing
problems
with $p^{th}$ power objectives, their analysis crucially uses the uniformity
of the exponent $p_e$
across all resources $e$. In this paper, we provide a clean extension of these
results to
heterogeneous exponents. In addition to obtaining a better (and tight)
competitive ratio than
previous results for heterogeneous exponents \cite{NagarajanW24}, our
algorithm and analysis are
simpler. We just use the natural greedy algorithm that allocates each request
so as to
(approximately) minimize the increase in objective. Our analysis is
elementary:
the main ingredient being the use of a $p$-subadditivity property.

We also consider the more general class of cost functions that exhibit
(dis)economies of scale,
which have been studied widely because they model energy consumption more
accurately than power cost
functions~\cite{AndrewsAZ16,AntoniadisIKMNP20,KrishnaswamyNPS14}.
Specifically, the cost of any
resource $e\in E$ under load $x$ is now given by $\sigma_e+\xi_e\cdot x^{p_e}$
if $x>0$
(and zero if $x=0$), where $\sigma_e>0$ corresponds to a start-up cost
(incurred whenever the
resource is turned on) and $\xi_e\cdot x^{p_e}$ corresponds to the
load-dependent part (as before).
We refer to these cost functions as DoS costs; note that these cost functions
are neither convex nor concave. Using our result on heterogeneous power
functions, in conjunction with a previously-known reduction from
\cite{NagarajanW24}, we also obtain an improved competitive
ratio for online network design under DoS costs.

\subsection{Problem Definition.}

\def\gnd{\ensuremath{\mathsf{GND}}\xspace}

We use the same framework as in \cite{EmekKLS20} and \cite{NagarajanW24},
referred to as
generalized network design (\gnd).
Let $E$ be a finite set of resources. Requests arrive online in the order
$1,2,\ldots,N$.
Each request $i\in [N]:=\{1,2,\cdots N\}$ has
\begin{enumerate}
  \item a feasible ``reply'' collection $\mathcal P_i\subseteq 2^E$ and
  \item a weight vector $w_i=(w_{i,e})_{e\in E}\in\R_{\ge 1}^E$.
\end{enumerate}
We note that the reply collection can be specified implicitly and the weights
are non-uniform.
If request $i$ is assigned reply $\pi_i\in\mathcal P_i$, then it contributes
$w_{i,e}$ units of
load to each resource $e\in \pi_i$. For any solution $\bpi
=(\pi_1,\ldots,\pi_N)$,
where $\pi_i\in {\cal P}_i$ for each $i\in [N]$, the induced load on resources
is given by
\[
  L_e(\bpi) := \sum_{i:e\in \pi_i} w_{i,e}, \qquad \forall e\in E .
\]
We consider two different objectives:
\begin{itemize}
  \item {\em Heterogeneous Power Functions.} Each resource $e\in E$ has a
  coefficient $\xi_e>0$ and
  an exponent $p_e\ge 1$. The objective is
  $\sum_{e\in E} h_e( L_e(\bpi)) $, where
  \[
    \refstepcounter{equation}
    \tag{\theequation}
    h_e(x) = \xi_e \cdot x^{p_e},\qquad \forall e\in E.
  \]
  \item {\em (Dis)economies of scale Functions.} Each resource $e\in E$ has a
  setup cost
  $\sigma_e>0$, coefficient $\xi_e>0$ and exponent $p_e\ge 1$.
  The DoS objective is $\sum_{e\in E} f_e (L_e(\bpi))$, where
  \[
    \refstepcounter{equation}
    \tag{\theequation}
    \label{eq:DoS}
    f_e(x) = \left\{
    \begin{array}{ll}
      0 & \mbox{ if } x=0 \\
      \sigma_e + \xi_e\cdot x^{p_e} & \mbox{ if } x>0
    \end{array}\right., \quad \forall e\in E.
  \]
  We also use the notation $q_e : = (\sigma_e/\xi_e)^{1/p_e}$, which is the
  value of load
  $x$ at which the two terms in function $f_e(x)$ become equal.
\end{itemize}
Let $p:= \max_{e\in E} p_e$ be the maximum exponent and $Q:=\max_{e\in E}
q_e$.
In this paper, we assume that $Q\ge 1$. Otherwise, the DoS cost function
reduces to heterogeneous
power costs with a factor $2$ loss.

\paragraph{Approximate reply oracles.}
We only assume that the single-request minimization problem can be solved {\em
approximately}.
This allows us to obtain polynomial-time online algorithms for a wide range of
network design
problems under DoS costs. Specifically, given any request $i$ and
``incremental costs''
$\{d_{e} \ge 0 : e\in E\}$ for each resource $e$, we assume that there is a
$\tau$-approximation
algorithm for the min-cost reply oracle:
\[
  \min_{\pi\in\mathcal P_i}\,\,\, \sum_{e\in \pi}d_{e}.
\]
This is the same oracle abstraction used previously in
\cite{EmekKLS20,NagarajanW24}.
We briefly mention below some applications of this framework
(these are also described in prior work).
\begin{itemize}
  \item Multicommodity routing: The resources are edges/nodes in a directed
  graph and each request
  $i$ is a source-destination pair $s_i,t_i$. The replies correspond to
  $s_i-t_i$ paths in the
  graph. The min-cost reply oracle is just shortest path: so $\tau=1$.
  \item Steiner connectivity: The resources are edges in an undirected graph
  and each
  request $i$ is a subset $T_i$ of nodes that need to get connected. The
  replies correspond to
  Steiner trees, and we have $\tau=1.39$ using an approximation algorithm for
  the Steiner tree
  problem~\cite{ByrkaGRS13}.
  \item Directed Steiner connectivity. The resources are edges/nodes in a
  directed graph and each
  request $i$ is a subset $T_i$ of nodes that need to get strongly connected.
  The replies
  correspond to strongly connected Steiner subgraphs. We have
  $\tau=O(\frac{\log^2 k}{\log \log k})$
  in quasi-polynomial time ($k=\max_i |T_i|$), using approximation algorithms
  for directed
  Steiner tree~\cite{GLL19,GhugeN20}.
\end{itemize}

\subsection{Results and Techniques.}
Our main result is for the online greedy algorithm that assigns each request
to
a $\tau$-approximately min-cost reply.
\begin{theorem}\label{thm:power-costs}
  The greedy algorithm for online generalized network design with
  heterogeneous power functions
  has competitive ratio $\left( \frac{1}{s^{1/p}-1}\right)^p\le
  \left(\frac{p}{\ln s}\right)^p$,
  where $p=\max_{e\in E}p_e$ and $s=1+\frac1\tau$.
\end{theorem}
When $\tau=1$ (an exact greedy algorithm), we obtain a ratio of $(1.443 \,
p)^p$ and when $\tau$
grows large the ratio tends to $(\tau \, p)^p$. This improves upon the
$2(\mathsf{e} \tau p)^p$
competitive ratio obtained previously in \cite{NagarajanW24}, where
$\mathsf{e}\approx 2.718$.
Our analysis is also simpler and completely elementary (the previous analysis
was based on a
convex dual-fitting argument). Our analysis exploits convexity and the
$p^{th}$ power function in
a number of ways, notably via coordinate-supermodular and $p$-subadditivity
properties. Our bound
for $\tau=1$ is optimal up to lower-order terms: \cite{Caragiannis08} gives a
lower bound of
$(1.443p)^p$ for deterministic online algorithms, even with uniform exponents.

In order to utilize this result for DoS costs, we need a more nuanced
guarantee, as in \cite{NagarajanW24}.
To this end, we use a slightly modified greedy algorithm and obtain:
\begin{theorem}\label{thm:power-nuanced}
  For any $\theta\in (1,s)$, there is an online algorithm for generalized
  network design with
  heterogeneous power functions that ensures
  \[
    \sum_{e\in E} \left(1- \theta^{-\frac1{p_e-1}}\right)^{p_e-1}\cdot h_e(
    L_e)\,\le\,
    \frac1{s-\theta} \cdot \OPT,
  \]
  where $s=1+\frac1\tau$, $\OPT = \sum_e h_e (O_e)$ and $L_e$ denotes the
  final load
  on each resource $e\in E$.
\end{theorem}

Roughly speaking, this result provides a stronger guarantee for resources with
smaller exponents.
The coefficient for resources $e$ with $p_e=1$ in the above expression equals
$1$ (which is the
limiting value as $p_e\rightarrow 1$).
The analysis for this result is almost identical to that for
Theorem~\ref{thm:power-costs}.
Combining Theorem~\ref{thm:power-nuanced} with the reduction from DoS
functions to
heterogeneous power functions, we obtain:
\begin{theorem}\label{thm:DoS}
  There is an online algorithm for generalized network design with DoS cost
  functions with
  competitive ratio at most
  $2\cdot \max\left\{ \frac{p}{\ln s} \cdot Q \,,\, \left(\frac{p}{\ln
  s}\right)^p\right\}$,
  where $p=\max_{e\in E}p_e$, $s=1+\frac1\tau$ and $Q = \max_{e\in E} q_e \geq
  1$.
\end{theorem}
This is an improvement over the $O\left(Q \tau + (\mathsf{e} p \tau)^p\right)$
competitive ratio
obtained in \cite{NagarajanW24}, which in turn was an improvement over the
offline $O\left(Q\tau+ 3^{p^2} \tau^p\right)$-approximation algorithm for \gnd
from \cite{EmekKLS20}.
When $\tau=1$, our competitive ratio is $O(p Q\,+\, (1.443 p)^p)$. This is
nearly tight: there is a
lower-bound of $\Omega(Q + (1.443p)^p)$ for any deterministic online algorithm
for \gnd with DoS costs~\cite{NagarajanW24}.

\subsection{Related Work.}
Online load balancing under {\em uniform} $p$ power objectives have been
extensively studied,
starting with the seminal work of \cite{AwerbuchAGKKV95}. This corresponds to
the special case where
requests are jobs, resources are machines and each ``reply'' corresponds to a
single machine
(that the job gets assigned to). The greedy algorithm was shown to obtain
a $\left(1.763 p +\Theta(\log p)\right)^p$-competitive ratio (assuming
$\tau=1$).
Later, \cite{Caragiannis08} obtained an improved bound of $\left(\frac{p}{\ln
2}\right)^p$ and
also showed that no deterministic online algorithm can achieve a better bound.
Both these results
relied crucially on uniform $p$, through the use of H\"older's inequality and
triangle inequality
for $\ell_p$-norms.

Some papers have considered load balancing in more general network design
problems.
For the {\em offline} setting, \cite{MakarychevS18} obtained a convex-program
based
$\left(\frac{p}{\ln (1+p)}\right)^p$-approximation algorithm for \gnd; this
approach does not work
in an online setting.
\cite{GuptaKP12} analyzed the online greedy algorithm for multicommodity
routing, which is a special
case of our setting, and obtained an $O(p)^p$ competitive ratio for uniform
exponent $p$. As mentioned before, \cite{NagarajanW24} obtained an
$2(\mathsf{e} \tau p)^p$ competitive ratio for \gnd with non-uniform $p_e$s.
Both these online
results use a dual-fitting approach on a convex relaxation for \gnd; the exact
dual formulations and
analysis are different.

The papers \cite{HarksHPV07,KlimmST19} also consider online \gnd with
non-uniform exponents $p_e$
and obtained a $(1.763 p)^{p+1}$ competitive ratio by analyzing the greedy
algorithm. However,
these results rely on {\em uniform weights} for each request, i.e.,
$w_{i,e}=w_i$ for all resources
$e\in E$. So, these results are incomparable to the earlier results on load
balancing
\cite{AwerbuchAGKKV95,Caragiannis08}. On one hand, the results
in~\cite{HarksHPV07,KlimmST19} are more general than
\cite{AwerbuchAGKKV95,Caragiannis08} because they handle general
resource allocation problems (\gnd). On the other hand, they are more
restrictive because they do
not handle the unrelated setting (where each job has a different weight on
each machine). Our result
is more general than all these prior results because it handles non-uniform
$p_e$s as well as
non-uniform weights $w_{i,e}$. We also improve the competitive ratio in
comparison
to \cite{HarksHPV07,KlimmST19}.

DoS cost functions have received considerable recent attention due to
applications in
energy-efficient computing. DoS functions are significantly harder than power
functions because
they are neither convex nor concave. Several
papers~\cite{AndrewsAZ16,AntoniadisIKMNP20,BansalGKNPS12,KrishnaswamyNPS14}
obtained
poly-logarithmic approximation/competitive ratios for multicommodity routing
on undirected graphs
in the special case of uniform DoS functions (across all resources $e$).
Subsequently, \cite{EmekKLS20} considered the much more general setting of
\gnd with non-uniform
DoS functions and obtained an approximation algorithm with ratio
$O\left(Q\tau+ 3^{p^2} \tau^p\right)$.
\cite{NagarajanW24} extended this to the online setting and also improved the
bound to
$O\left(Q\tau+(\mathsf{e} \tau p)^p\right)$. We improve this bound further as
stated in Theorem~\ref{thm:DoS}.
We note that one cannot hope to eliminate the
polynomial dependence on $Q$ in these (non-uniform) \gnd results even in the
offline setting: this
is due to the hardness of approximation for directed Steiner
forest~\cite{DodisK99}.

There have also been several results on (more general) online covering
problems under convex
objectives, see e.g., \cite{GuptaKP12,
AnandGK12,
DevanurH18,
AzarBCCCGHKNNP16,
NagarajanS17,
HuangK19,
KesselheimMPS25}.
Most of these papers build on the online primal-dual approach: our technique
is very different
(and more elementary). We note that our idea of utilizing the $p$-subadditive
property in the
analysis was partly motivated by the integral online algorithms in
\cite{KesselheimMPS25}.

\section{The Generic Greedy Algorithm.}
Let $m:= |E|$. We consider a separable convex objective
$\Phi:\R_+^m\rightarrow \R_+$ of the form:
\[
  \Phi(\bx) = \sum_{e\in E} \varphi_e(x_e),
\]
where the functions $\varphi_e$ satisfy the following.
\begin{assumption}
  \label{assmp:function}
  For each $e\in E$, function $\varphi_e:\R_+\rightarrow\R_+$ has the form
  $\varphi_e(x)=c_e\cdot x^{p_e}$ where $c_e>0$ and $p_e\ge 1$. We let
  $p=\max_{e\in E} p_e$ denote the maximal exponent.
\end{assumption}

We will later use $\Phi$ as the original objective function in order to
analyze the natural greedy
algorithm (Theorem~\ref{thm:power-costs}). We will also use $\Phi$ as a
modified
objective in order to obtain a nuanced bound (Theorem~\ref{thm:power-nuanced})
that
turns out to be crucial in handling DoS costs (Theorem~\ref{thm:DoS}).

\paragraph{Useful properties of $\Phi$.} We now state some useful properties
of $\Phi$ that are
based on convexity and the $p^{th}$ power form.
\begin{lemma}[Scaling]
  \label{lem:scaling}
  For any $\rho\ge 1$ and $\bx\in \R_+^m$, we have $\Phi(\rho\bx)\le
  \rho^p\cdot \Phi(\bx)$.
\end{lemma}
\begin{proof}
  By Assumption~\ref{assmp:function},
  for each $e\in E$ and $x\ge 0$ we have
  $\varphi_e(\rho\cdot x)= \rho^{p_e}\cdot \varphi_e(x)\le \rho^p\cdot
  \varphi_e(x)$, where we used
  $\rho\ge 1$. So, we have
  $\Phi(\rho \cdot \bx) = \sum_{e\in E} \varphi_e(\rho\cdot x_e)\le
  \rho^{p}\cdot \sum_{e\in E}
  \varphi_e( x_e)= \rho^{p}\cdot\Phi(\bx)$.
\end{proof}

\begin{lemma}[$p$-subadditive]\label{lem:p-subadd}
  For any $\bx,\by\in \R^m_+$, we have $\Phi(\bx+\by)^{1/p}\le
  \Phi(\bx)^{1/p}+\Phi(\by)^{1/p}$.
\end{lemma}
\begin{proof}
  If $\Phi(\bx)=0$ then we must have $\bx=\bzero$, in which case the lemma is
  trivially true. (The
  same argument holds when $\Phi(\by)=0$.) We now assume that both $\Phi(\bx),
  \Phi(\by) >0$.
  For any $\lambda \in (0,1)$, by convexity of $\Phi$, we have:
  \[
    \Phi(\bx+\by)
    = \Phi\!\left(\lambda\cdot \frac{\bx}{\lambda} +
    (1-\lambda)\cdot\frac{\by}{1-\lambda}\right)
    \leq \lambda\, \Phi\!\left(\frac{\bx}{\lambda}\right) + (1-\lambda)\,
    \Phi\!\left(\frac{\by}{1-\lambda}\right)
    \leq \lambda^{1-p} \, \Phi(\bx) + (1-\lambda)^{1-p}\, \Phi(\by).
  \]
  The last inequality uses Lemma~\ref{lem:scaling}.
  Now, choosing the value $\lambda = \frac{\Phi(\bx)^{1/p}}{\Phi(\bx)^{1/p} +
  \Phi(\by)^{1/p}}$
  makes the right-hand side equal to $\bigl(\Phi(\bx)^{1/p} +
  \Phi(\by)^{1/p}\bigr)^p$, which yields
  the desired result.
\end{proof}

\begin{lemma}[Monotone increments]\label{lem:mon-increment}
  For any $e\in E$ and values $x'\ge x\ge 0$ and $v\in \R_+$,
  \[
    \varphi_e(x+v) - \varphi_e(x) \le \varphi_e(x'+v) - \varphi_e(x').
  \]
\end{lemma}

\begin{proof}
  As $\varphi_e:\R_+\rightarrow \R_+$ is convex, for any $v\ge 0$
  the univariate difference $t \mapsto \varphi_e(t+v) - \varphi_e(t)$ is
  non-decreasing in $t$.
  As $x \leq x'$, we get $\varphi_e(x+v ) - \varphi_e(x ) \leq \varphi_e(x' +v
  ) - \varphi_e(x' )$.
\end{proof}

\begin{lemma}[Coordinate-wise superadditivity]\label{lem:coord-supadd}
  For any $e \in E$, $a\ge 0$ and $t_1,\dots,t_k\ge 0$, we have
  \[
    \sum_{j=1}^{k} \left[ \varphi_e(a +t_j) - \varphi_e(a)\right] \quad
    \leq\quad
    \varphi_e\!\left(a+\sum_{j=1}^{k} t_j\right) - \varphi_e(a).
  \]
\end{lemma}
\begin{proof}
  Define function $g_e(x) := \varphi_e(a +x) - \varphi_e(a)$ for $x\ge 0$.
  Clearly, $g_e$ is
  convex with $g_e(0)=0$.
  Let $T:=\sum_{j=1}^k t_j$. If $T = 0$, then every $t_j = 0$ and both sides
  vanish; so
  assume $T > 0$. By convexity, we have for each $j\in [k]$:
  \[
    g_e(t_j) \,=\, g_e\left( \frac{t_j}{T}\cdot T + (1-\frac{t_j}{T}) \cdot
    0\right) \;\leq\;
    \frac{t_j}{T}\, g_e(T) + (1-\frac{t_j}{T}) \, g_e(0)
    \;=\; \frac{t_j}{T}\, g_e(T).
  \]
  Adding over $j\in [k]$ gives
  $\sum_{j=1}^k g_e(t_j) \le g_e(T)\cdot \sum_{j=1}^k \frac{t_j}{T} = g_e(T)$,
  which completes the proof.
\end{proof}

\paragraph{The $\Phi$ greedy algorithm.}
We consider the greedy algorithm that assigns each online request $i\in [N]$
to a reply
$\pi_i\in {\cal P}_i$ that (approximately) minimizes the increase in the
$\Phi$ objective value.
Formally, when request $i\in [N]$ arrives let $\ell_e$ be the current online
load on each resource
$e\in E$. The marginal increase in the objective from assigning request $i$ to
resource $e$ is
\[
  d_{i,e}(\bell) := \varphi_e \left( \ell_e+w_{i,e} \right)- \varphi_e \left(
  \ell_e \right).
\]
Then, the online algorithm runs the min-cost reply oracle on costs
$\{d_{i,e}(\ell) : e\in E\}$ to
obtain $\pi_i\in {\cal P}_i$ such that
\[
  \sum_{e\in \pi_i}d_{i,e}(\bell) \quad \le\quad \tau\cdot
  \min_{\pi\in\mathcal P_i}
  \sum_{e\in \pi}d_{i,e}(\bell).
\]
Below, we use $s=1+\frac1\tau \in (1,2]$ by $\tau \ge 1$.
We let $L_e$ denote the {\em final} online load on each resource $e\in E$.
Then the online
algorithm's objective is $\ALG := \Phi(\bL)=\sum_{e\in E}\varphi_e(L_e)$.

For the analysis, let $\pi^*_i\in\mathcal P_i$ denote the reply used by the
{\em offline optimum}
for each request $i\in [N]$. Let $O_e:=\sum_{i:e\in \pi^*_i}w_{i,e}$ denote
the final optimal load
on each resource $e\in E$.

\begin{lemma}[Master lemma]\label{lem:master}
  Letting $\bL$ and $\bO$ denote the load vectors of the algorithm and the
  optimum, we have
  \[
    s \cdot \Phi(\bL) \;\leq\; \Phi(\bL + \bO),
    \qquad \text{where } s = \tfrac{1+\tau}{\tau}.
  \]
\end{lemma}

\begin{proof}
  Let $\bell^{(i)}\in \R^m_+$ denote the algorithm's load vector immediately
  after processing
  request $i\in [N]$. So, $\bell^{(0)}=\bzero$ and $\bell^{(N)}=\bL$. Then,
  the increase in the
  $\Phi$ objective when request $i$ is processed is
  $\Delta_i := \Phi(\bell^{(i)}) - \Phi(\bell^{(i-1)})$. By a telescoping sum
  and using
  $\Phi(\bzero)=0$,
  \[
    \Phi(\bL) \;=\; \sum_{i=1}^{N} \Delta_i .
  \]
  For each $i\in [N]$, we have
  \[
    \begin{aligned}
      \Delta_i & \;=\; \sum_{e \in \pi_i} \bigl[\varphi_e(\ell^{(i-1)}_e +
      w_{i,e}) -
      \varphi_e(\ell^{(i-1)}_e)\bigr]
      \;=\; \sum_{e\in\pi_i} d_{i,e}\bigl(\bell^{(i-1)}\bigr) \le
      \tau \sum_{e \in \pi_i^\ast} d_{i,e}\bigl(\bell^{(i-1)}\bigr) \\
      & = \tau \sum_{e \in \pi_i^\ast} \bigl[\varphi_e(\ell^{(i-1)}_e +
      w_{i,e}) -
      \varphi_e(\ell^{(i-1)}_e)\bigr]\,\le\, \tau \sum_{e \in \pi_i^\ast}
      \bigl[\varphi_e(L_e + w_{i,e}) - \varphi_e(L_e)\bigr] .
    \end{aligned}
  \]
  The first inequality is by the $\tau$-approximate greedy choice and the
  second inequality is by
  Lemma~\ref{lem:mon-increment}
  where we use $\bell^{(i-1)}\le \bL$.
  Adding over all requests $i$,
  \refstepcounter{equation}\label{eq:main-lem-1}
  \[
    \tag*{\raisebox{-11.811pt}{(\theequation)}\kern 0.00307pt}
    \begin{aligned}
      \Phi(\bL) & \, \le \, \tau \sum_{i=1}^N \sum_{e \in \pi_i^\ast}
      \bigl[\varphi_e(L_e + w_{i,e})
      - \varphi_e(L_e)\bigr] \,=\, \tau \sum_{e \in E} \, \sum_{i : e\in
      \pi^*_i}
      \bigl[\varphi_e(L_e + w_{i,e}) - \varphi_e(L_e)\bigr] \\
      & \,\le\, \tau \sum_{e \in E} \left[\varphi_e\left(L_e + \sum_{i : e\in
      \pi^*_i}
      w_{i,e}\right) - \varphi_e(L_e)\right] \,=\, \tau
      \sum_{e \in E} \left[\varphi_e(L_e +O_e) - \varphi_e(L_e)\right] \\
      & \,=\, \tau \cdot \Phi(\bL+\bO) - \tau \cdot \Phi(\bL).
    \end{aligned}
    \addtolength{\belowdisplayskip}{0.308pt}
    \addtolength{\belowdisplayshortskip}{0.308pt}
  \]
  The inequality in \eqref{eq:main-lem-1} is by Lemma~\ref{lem:coord-supadd}
  applied to each resource $e\in E$.
  Finally, rearranging the last expression, we get
  $(\tau+1)\cdot \Phi(\bL) \le \tau \cdot \Phi(\bL+\bO)$, which completes the
  proof.
\end{proof}

\paragraph{Competitive ratio of the natural greedy algorithm.}
Here we consider the greedy algorithm applied with the original objective in
\gnd,
i.e., $\varphi_e(x) = h_e( x) = \xi_e\cdot x^{p_e}$. We prove the competitive
ratio stated in
Theorem~\ref{thm:power-costs}.
Indeed, using Lemma~\ref{lem:master}, we obtain
$s \cdot \Phi(\bL) \;\leq\; \Phi(\bL + \bO) $.
Moreover, by Lemma~\ref{lem:p-subadd},
$ \Phi(\bL+\bO)^{1/p} \;\leq\; \Phi(\bL)^{1/p} + \Phi(\bO)^{1/p}$. Combining
these two inequalities,
\[
  s^{1/p}\, \Phi(\bL)^{1/p} \;\leq\; \Phi(\bL)^{1/p} + \Phi(\bO)^{1/p},
  \qquad\text{i.e.,}\qquad
  \bigl(s^{1/p} - 1\bigr)\, \Phi(\bL)^{1/p} \;\leq\; \Phi(\bO)^{1/p}.
\]
Raising both sides to the $p$-th power completes the proof of
Theorem~\ref{thm:power-costs}.

\subsection{Modified Greedy with Resource-Dependent Bound.}
We now prove Theorem~\ref{thm:power-nuanced}.
The reason for this refined bound is that the ``DoS to power-functions''
reduction from \cite{NagarajanW24}
(see Lemma~\ref{lem:fn-apx}) loses a factor $Q$ only on the linear terms while
the non-linear terms
suffer only a constant factor. Theorem~\ref{thm:power-nuanced} allows us to
bound
these two terms separately with different coefficients, which is crucial for
obtaining the overall
DoS bound in Theorem~\ref{thm:DoS}.

We now consider a modified greedy algorithm that enables proving the nuanced
resource-dependent
bound in Theorem~\ref{thm:power-nuanced}. To this end, we work with a
different
objective function $\Phi$ described below.
Fix a parameter $\theta \in (1, s)$ where $s=\frac{1+\tau}{\tau}$ as before.
Define per-resource weights as:
\[
  r_e \;:=\;
  \begin{cases}
    \left(1 - \theta^{-\frac1{p_e - 1}}\right)^{p_e - 1} & \text{if } p_e > 1,
    \\[2pt]
    1 & \text{if } p_e = 1 ,
  \end{cases}
\]
Then, the objective function is
\[
  \refstepcounter{equation}
  \tag{\theequation}
  \label{eq:new-Phi}
  \Phi(\bx) \;:=\; \sum_{e\in E} r_e \cdot h_e(x_e) = \sum_{e\in E} r_e \cdot
  \xi_e \cdot x_e^{p_e}.
\]
Note that the new functions $\varphi_e(x) = r_e \cdot \xi_e \cdot x^{p_e}$
satisfy
Assumption~\ref{assmp:function}.
The algorithm runs $\tau$-approximate greedy with this modified objective
$\Phi$.

\begin{lemma}[Nuanced bound]\label{lem:nuanced}
  Letting $\bL$ and $\bO$ denote the load vectors of the algorithm and the
  optimum, we have
  \[
    \sum_{e \in E} r_e\cdot h_e(L_e)
    \;\leq\;
    \frac{1}{s - \theta} \cdot \sum_{e \in E} h_e(O_e),
  \]
  where $s=\frac{1+\tau}{\tau}$.
\end{lemma}
\begin{proof}
  Applying Lemma~\ref{lem:master} to the new
  objective from \eqref{eq:new-Phi} gives
  \[
    \refstepcounter{equation}
    \tag{\theequation}
    \label{eq:master-theta}
    s\cdot \Phi (\bL) \;\leq\; \Phi (\bL + \bO) \;=\; \sum_{e\in E} r_e \cdot
    h_e(L_e + O_e).
  \]
  We will show that
  \[
    \refstepcounter{equation}
    \tag{\theequation}
    \label{eq:mix-e}
    h_e(L_e + O_e) \;\leq\; \theta\cdot h_e(L_e) + \frac1{r_e}\cdot
    h_e(O_e),\qquad \forall e\in E.
  \]
  First, we complete the proof of the lemma assuming \eqref{eq:mix-e}.
  Indeed, we then get
  \[
    s\cdot \Phi (\bL) \;\leq\; \sum_{e\in E} r_e \cdot h_e(L_e + O_e) \,\le\,
    \sum_{e\in E} r_e \cdot \theta\cdot h_e(L_e) + \sum_{e\in E} h_e(O_e)
    \,=\,
    \theta\cdot \Phi(\bL) + \sum_{e\in E} h_e(O_e).
  \]
  Collecting the $\Phi(\bL)$ terms, we obtain $\Phi(\bL)\le
  \frac{1}{s - \theta} \cdot \sum_{e \in E} h_e(O_e)$ as desired.

  It remains to prove \eqref{eq:mix-e}. For any $e$ with $p_e=1$, we have
  $r_e=1$
  and so
  \[
    h_e(L_e + O_e)=h_e(L_e)+h_e(O_e)\le \theta\cdot h_e(L_e) +
    \frac1{r_e}\cdot h_e(O_e),
  \]
  where we used $\theta>1$.

  We now consider $e\in E$ with $p_e>1$: so $r_e=\left(1 - \theta^{-\frac1{p_e
  - 1}}\right)^{p_e - 1}$.
  Let $\lambda_e := \theta^{-\frac1{p_e-1}}$. Note that $\lambda_e\in (0,1)$
  and
  \[
    \lambda_e^{\,1-p_e} = \theta
    \qquad \mbox{and} \qquad
    (1-\lambda_e)^{\,1-p_e} = \frac1{r_e}.
  \]
  We now use a ``mixing inequality''
  (see Lemma~\ref{lem:mixing} below)
  to function $h_e$ with exponent $q=p_e$ and values $x = L_e$, $y = O_e$,
  and $\lambda_e\in (0,1)$ to get:
  \[
    h_e(L_e+O_e)\, \le\, \lambda_e^{1-p_e} \cdot h_e(L_e) +
    (1-\lambda_e)^{1-p_e}\cdot h_e(O_e)
    \,=\, \theta \cdot h_e(L_e) + \frac{1}{r_e}\cdot h_e(O_e),
  \]
  which proves \eqref{eq:mix-e}.
\end{proof}

We now state and prove the mixing inequality used in the proof above.

\begin{lemma}[Mixing inequality]\label{lem:mixing}
  Consider the 1-dimensional function $g(x):= c\cdot x^q$ where $c>0$ and
  $q\ge 1$.
  Then, for any $x,y\ge 0$ and $\lambda \in (0,1)$, we have
  \[
    g(x+y) \quad \leq\quad \lambda^{1-q}\cdot g(x) \;+\;
    (1-\lambda)^{1-q}\cdot g(y).
  \]
\end{lemma}
\begin{proof}
  Observe that $g$ is a convex function. So,
  \[
    g(x+y) \le \lambda\cdot g\left(\frac{x}\lambda\right) \,+\,
    (1-\lambda)\cdot
    g\left(\frac{y}{1-\lambda}\right) = \lambda^{1-q}\cdot g(x) \,+\,
    (1-\lambda)^{1-q}\cdot g(y) ,
  \]
  which completes the proof.
\end{proof}

Theorem~\ref{thm:power-nuanced} now follows directly from
Lemma~\ref{lem:nuanced} and the definition of coefficients $r_e$.

\section{Application to DoS Costs.}
We now prove Theorem~\ref{thm:DoS} for DoS costs.
We will combine Theorem~\ref{thm:power-nuanced} with a known ``reduction''
from DoS cost functions~\eqref{eq:DoS} to power functions. Specifically,
Lemma~1
in \cite{NagarajanW24} states:
\begin{lemma}[\cite{NagarajanW24}]\label{lem:fn-apx}
  Given any DoS cost function $f_e(x)$ of the form~\eqref{eq:DoS},
  define two power functions
  $h_{e_1}(x) \,:= \,\xi_e q_e^{p_e-1}\cdot x $ and $h_{e_a}(x) \,:=
  \xi_e\cdot x^{p_e}$
  where $q_e = (\sigma_e/\xi_e)^{1/p_e}$. Then, for any $x\in \{0\}\cup
  \R_{\ge 1}$,
  \[
    \frac12 \cdot \left( h_{e_1}(x) + h_{e_a}(x)\right) \,\le\, f_e(x)\le
    \max\{q_e,1\}
    \cdot h_{e_1}(x) + h_{e_a}(x) \le \max\{q_e,1\} \cdot \left( h_{e_1}(x) +
    h_{e_a}(x)\right) .
  \]
\end{lemma}

This reduces DoS costs to power functions in a pointwise manner, at the loss
of factor $2q_e$.
Moreover, it provides a nuanced bound where only the linear term $h_{e_1}(x)$
incurs the (large)
factor $2q_e$ loss and the non-linear term $h_{e_a}(x)$ incurs just a factor
$2$ loss. As a
result, one can utilize a nuanced competitive ratio for power functions, such
as
Theorem~\ref{thm:power-nuanced}, to obtain a better overall competitive ratio
for DoS
costs. This was also the approach in \cite{NagarajanW24} (see Section~5),
where they relied on a
weaker competitive ratio for power functions.
We now show how Theorem~\ref{thm:power-nuanced} can be used to obtain the
improved
competitive ratio for DoS stated in Theorem~\ref{thm:DoS}.

\def\I{\ensuremath{{\cal I}}\xspace}
\def\J{\ensuremath{{\cal J}}\xspace}
\paragraph{Reduction from DoS to power cost functions.}
For completeness, we restate the reduction from \cite{NagarajanW24}.
Given an instance \I of \gnd with DoS cost functions $\{f_e\}_{e\in E}$ as in
\eqref{eq:DoS},
we use Lemma~\ref{lem:fn-apx} to define a new instance \J
of \gnd with {\em power functions}, as follows. For each original resource
$e\in E$, we have two
resources $e_1$ and $e_a$. Let $E_1:=\{e_1:e\in E\}$ and $E_a:=\{e_a:e\in
E\}$: so the resources in
\J are $E'=E_1\sqcup E_a$. Define scalars $\xi_{e_1}:=\xi_e q_e^{p_e-1}$ and
$\xi_{e_a}:=\xi_e$ for
all $e\in E$. Also, define exponents $p_{e_1}:=1$ and $p_{e_a}:=p_e$ for all
$e\in E$. The power
functions in instance \J are $g_{r}(x):= \xi_{r}\cdot x^{p_{r}}$ for all
resources $r\in E'$. The
reply-collections are extended so that for each reply $\pi\in {\cal P}_i$ in
\I, there is a
corresponding reply in \J that contains {\em both} copies $\{e_1,e_a\}$ of all
resources $e\in \pi$.
We now apply the online algorithm in Theorem~\ref{thm:power-nuanced} for
heterogeneous
power functions to instance \J, and use this solution for the DoS cost
instance \I.

Let $\bL=(L_e:e\in E')$ denote the final loads in the online algorithm. We
separate the linear and
non-linear terms in instance $\J$ as follows:
\[
  T_1 := \sum_{e\in E_1} h_e(L_e) \quad \mbox{and}\quad T_a:= \sum_{e\in E_a}
  h_e(L_e).
\]
Using Theorem~\ref{thm:power-nuanced}, we obtain:
\[
  \begin{aligned}
    \frac{1}{s-\theta}\cdot \OPT_{\J} & \,\ge\, \sum_{e\in E'} \left(1-
    \theta^{-\frac1{p_e-1}}\right)^{p_e-1}\cdot h_e( L_e) \,=\, T_1 +
    \sum_{e\in E_a} \left(1-
    \theta^{-\frac1{p_e-1}}\right)^{p_e-1}\cdot h_e( L_e) \\
    & \,\ge\, T_1+\left(1-\theta^{-\frac1{p-1}}\right)^{p-1}\cdot
    \sum_{e\in E_a} h_e( L_e) \,=\,
    T_1+ \left(1-\theta^{-\frac1{p-1}}\right)^{p-1}\cdot T_a .
  \end{aligned}
\]
The second inequality above uses the fact that the function $ \bigl(1 -
\theta^{-1/r}\bigr)^{r}$ is
non-increasing for $r>0$ given that $\theta \in (1,s)$. Now, using $\OPT_{\J}
\le 2\cdot \OPT_{\I}$
(which follows from Lemma~\ref{lem:fn-apx}), we obtain:
\[
  \refstepcounter{equation}
  \tag{\theequation}
  \label{eq:separate-dos-bounds}
  T_1\le
  \frac{2}{s-\theta}\cdot \OPT_{\I}\quad\mbox{and}\quad T_a \le
  \frac{2}{s-\theta}\cdot \left(1-\theta^{-\frac1{p-1}}\right)^{-p+1}\cdot
  \OPT_{\I}
\]
Using Lemma~\ref{lem:fn-apx}, we can bound the DoS objective
of the online solution as follows:
\[
  \begin{aligned}
    \sum_{e\in E} f_e(L_e) & \le \sum_{e\in E_1} \max\{q_e,1\} \cdot h_e(L_e)
    + \sum_{e\in E_a} h_e(L_e) \\
    & \le Q\cdot T_1+T_a \le \frac{2}{s-\theta}\cdot \max\left\{ Q \,,\,
    \left(1-\theta^{-\frac1{p-1}}\right)^{-p+1}\right\}\cdot \OPT_{\I},
  \end{aligned}
\]
where $Q = \max_{e\in E} q_e$. (Recall that we assume $Q\ge 1$:
otherwise Lemma~\ref{lem:fn-apx} with
Theorem~\ref{thm:power-costs} directly implies an $O(\left(\frac{p}{\ln
s}\right)^p)$
competitive ratio for DoS costs.).

Setting $\theta=s^{1-1/p}<s$, we have:
\[
  \frac{1}{s-\theta} = \frac{1}{s^{1-1/p} \left(s^{1/p}-1\right)}\le \frac{1}{
  s^{1/p}-1 }\le
  \frac{p}{\ln s} ,\,\,\mbox{ and}
\]
\[
  \frac{1}{s-\theta} \cdot \left(1-\theta^{-\frac1{p-1}}\right)^{-p+1} =
  \left( \frac{1}{s^{1/p}-1}\right)^p \le \left(\frac{p}{\ln s}\right)^p.
\]
Hence, we obtain a competitive ratio at most:
\[
  2\cdot \max\left\{ \frac{p}{\ln s} \cdot Q \,,\, \left(\frac{p}{\ln
  s}\right)^p\right\} .
\]
This completes the proof of Theorem~\ref{thm:DoS}.

\section*{Acknowledgments.}
We used the AI tool Claude Opus 5 in coming up with the specific form of the
re-weighting function
in Theorem~\ref{thm:power-nuanced}.
Specifically, we provided Lemmas~\ref{lem:master}
and \ref{lem:mixing} as input
facts, and asked the AI tool for the best resource-dependent coefficients in
the left-hand-side of
Theorem~\ref{thm:power-nuanced} such that resources with $p_e=1$ have
coefficient one.
We verified the details and wrote the proof of
Lemma~\ref{lem:nuanced} ourselves.
The authors take full responsibility for the correctness of the results.

\bibliographystyle{alpha}
{\small
\bibliography{refs}
}
\end{document}